\documentclass[prl,amssymb,twocolumn,showpacs,superscriptaddress,floatfix]{revtex4}

\usepackage{graphicx}
\usepackage{bm}
\usepackage{mathptmx}

\begin{document}

\title{Strong Dynamical Heterogeneities in the Violation of the Fluctuation-Dissipation Theorem in Spin Glasses}

\author{F. Rom\'a}
\affiliation{Centro At{\'{o}}mico Bariloche, 8400 San Carlos de
Bariloche, R\'{\i}o Negro, Argentina} \affiliation{Departamento de
F\'{\i}sica, Universidad Nacional de San Luis, 5700 San Luis,
Argentina}
\author{S. Bustingorry}
\affiliation{Centro At{\'{o}}mico Bariloche, 8400 San Carlos de
Bariloche, R\'{\i}o Negro, Argentina}
\author{P. M. Gleiser}
\affiliation{Centro At{\'{o}}mico Bariloche, 8400 San Carlos de
Bariloche, R\'{\i}o Negro, Argentina}
\author{D. Dom\'{\i}nguez}
\affiliation{Centro At{\'{o}}mico Bariloche, 8400 San Carlos de
Bariloche, R\'{\i}o Negro, Argentina}

\begin{abstract}
We analyze numerically the violation of the
fluctuation-dissipation theorem (FDT) in the $\pm J$
Edwards-Anderson (EA) spin glass model. Using single spin
probability densities we reveal the presence of strong dynamical
heterogeneities, which correlate with ground state information.
The physical interpretation of the results shows that the spins in
the EA model can be divided in two sets. In 3D, one set forms a
compact structure which presents a coarsening-like behavior with
its characteristic violation of the FDT, while the other
asymptotically follows the FDT. Finally, we compare the dynamical
behavior observed in 3D with 2D.
\end{abstract}

 %Numerical simulation studies

\date{\today}

\maketitle

The study of glassy behavior, characterized by out-of-equilibrium
dynamics that present very long relaxation times, involves a vast
range of different systems such as spin glasses \cite{Mezard},
structural glasses, colloidal and polymeric systems, and granular
systems to name just a few \cite{lesuch1}. These systems present
out-of-equilibrium properties such as aging and violation of the
fluctuation-dissipation theorem (FDT) \cite{Crisanti,lesuch2},
whose characterization  can be used as signatures of typical
dynamical behaviors. For example, according to their deviation
from the FDT they can be classified into three different groups:
coarsening, structural glass and spin glass systems
\cite{Crisanti}. These cases have been theoretically described in
terms of replica-symmetry breaking (RSB): in coarsening systems
replica symmetry is unbroken, structural glasses correspond to
one-step RSB and spin glasses to full RSB \cite{Crisanti}. In the
case of short-range spin glasses, most results on the violation of
the FDT have been qualitatively and quantitatively described in
the framework of mean field theory (full RSB)
\cite{Crisanti,Gaviro}. Also direct experimental evidence on the
violation of the FDT in an insulating spin glass \cite{Ocio} have
been fitted by mean field theory.

In this work we show that, in contrast to these results,  the
violation of the FDT in the 3D $\pm J$ Edwards-Anderson (EA) model
is the result of two components with completely different
behaviors: one that tends to satisfy the FDT relation, and another
which presents a violation of this relation similar to coarsening
systems. The behavior of the latter component seems to be
compatible with the droplet picture scenario \cite{Fisher}.

We reveal the presence of two components in the violation of the
FDT by a careful analysis of dynamical heterogeneities.  In
principle, two different approaches may be used to study the
development of dynamical heterogeneities. On one hand, it is
possible to use space or time \textit{coarse-grained} protocols
\cite{Castillo03,Chamon04}. On the other hand, it is possible to
use single spin observables looking for a direct observation of
the local heterogeneities. In any of these approaches, if one is
interested in making disorder averages and chooses to identify the
spins by their position in the lattice, trivial results are
obtained, since the difference between spins are washed out
\cite{Montanari1}. Here we take a local approach, but in contrast
with previous works, we properly include disorder averages by
using a constrained structure of the ground state called {\em
backbone} (see below), which allows for much insight in the
analysis of dynamical heterogeneities \cite{Roma2}.

We begin our analysis by using single spin probability densities
to show that dynamical heterogeneities are present in the
violation of FDT in the 3D $\pm J$ EA model. These heterogeneities
present a non-trivial complex structure with bimodal distributions
\cite{Ricci,Jung,Roma2}, and we refer them as {\em strong}, in
contrast to {\em weak} dynamical heterogeneities which present
unimodal distributions with an elongated shape. Then, as done in
2D in \cite{Roma2}, we use information from the ground state
topology to establish a quantitative relation between spatial and
dynamical heterogeneities. This analysis allows us to present a
new physical interpretation of the violation of FDT. Finally we
compare the violation of FDT in 2D and 3D, which reveals
fundamental differences.

We consider the EA model for spin glasses \cite{Mezard}, defined
on a $D$-dimensional lattice of linear size $L$ with periodic
boundary conditions. The Hamiltonian of the model is
\begin{equation}
H = \sum_{\langle i,j \rangle} J_{ij} \sigma_{i} \sigma_{j}
\end{equation}
where $\sigma_i = \pm 1$ is the spin variable and $\langle i,j
\rangle$ indicates a sum over the $2D$ nearest neighbors. The
coupling constants $J_{ij} = \pm J$ are random variables chosen
from a symmetric bimodal distribution. The time evolution of the
model is governed by a standard Glauber dynamics with sequential
random updates using a continuous time Monte Carlo algorithm
\cite{Bortz}.

In order to investigate the violation of the FDT we use a
two-times protocol which emphasizes the out-of-equilibrium
character of the dynamics. The initial condition corresponds to a
quench from $T=\infty$ to the temperature of interest $T<T_c =
1.12$ \cite{Katz} at $t=0$, where each spin takes a random value
$\sigma_i = \pm 1$. From this initial condition different
two-times quantities are analyzed, which depend on both the
waiting time $t_w$, when the measurement begins, and a given time
$t>t_w$ \cite{Crisanti,lesuch2}.

We focus on the single spin two-times correlation function,
defined as ${\cal C}_i \equiv \langle C_i(t,t_w) \rangle = \langle
\sigma_i(t) \sigma_i(t_w) \rangle$ , with $t>t_w$, and where
$\langle \dots \rangle$ is an average over thermal histories, that
is over different initial conditions and realization of the
thermal noise. In terms of the single spin correlation, the global
two-times correlation function can be expressed as
\begin{equation}
C = \left [ \frac{1}{N} \sum_{i=1}^N {\cal C}_i \right ]_{av} =
\left [ \frac{1}{N} \left \langle \sum_{i=1}^N \sigma_i(t)
\sigma_i(t_w) \right \rangle \right ]_{av} \label{corr}
\end{equation}
where $N =L^D$ and $[...]_{av} $ indicates average over different
realizations of bond disorder (samples). In the same way the
global integrated response function is defined in term of the
single spin integrated response function as
\begin{equation}
\chi = \left [ \frac{1}{N} \sum_{i=1}^N \chi_i \right ]_{av} =
\left [\frac{1}{Nh} \left \langle \sum_{i=1}^N \sigma_i(t)
sign(h_i(t_w)) \right \rangle \right ]_{av} \label{magg}
\end{equation}
where $\chi_i \equiv \langle \sigma_i(t) sign(h_i(t_w))/h \rangle$
and, as usually, a random field of intensity $h$ is switched on at
time $t_w$ \cite{Crisanti,lesuch2}. All the results presented
correspond to $T=0.8$, $h=0.1$ and $10^4$ thermal histories for
each sample.

\begin{figure}
\includegraphics[width=\linewidth,clip=true]{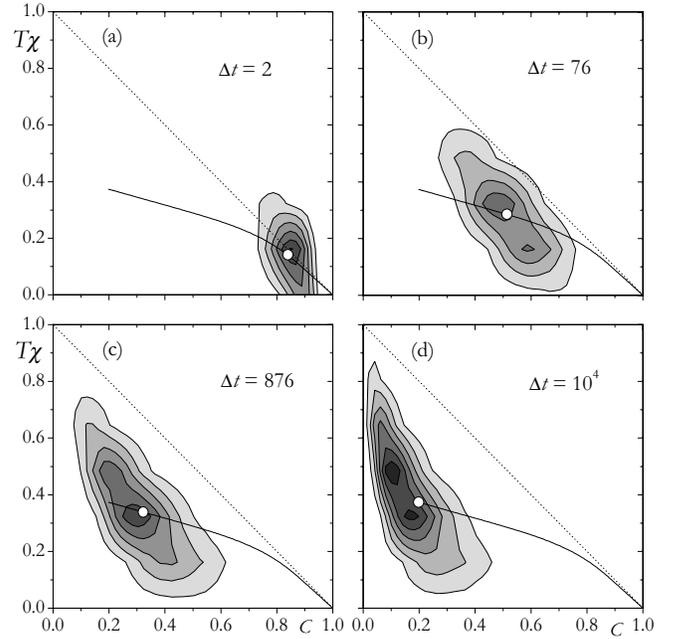}
\caption{\label{figure1} Projection of the JPD of single spin
correlations and integrated responses, $\rho[{\cal
C}_i(t,t_w),T\chi_i(t,t_w)]$, for $L=20$ and $t_w=10$. The
parametric plot of the global quantities, $T\chi(C)$, is
reproduced in all the figures as a continuous line. The white dot
indicates the mean value of the distribution for each $\Delta t$.}
\end{figure}

Using the global quantities defined in Eqs. (\ref{corr}) and
(\ref{magg}), the out-of-equilibrium fluctuation-dissipation relation
can be written as
\begin{equation}
T \chi(t,t_w)=X(C)\left[ 1-C(t,t_w) \right] \label{fdt}
\end{equation}
where $X$ is the fluctuation-dissipation ratio (FDR)
\cite{Peliti}. A useful representation of Eq.~(\ref{fdt}) is the
parametric plot of $T\chi$ vs. $C$ \cite{Crisanti,lesuch2}. When
the FDT holds the FDR is $X=1$ and the parametric plot shows a
linear relation with unitary slope. In an out-of-equilibrium
situation the FDT does not longer hold and two regimes are
observed: for $t/t_w \ll 1$ the system shows quasi-equilibrium
with $X=1$, while for $t/t_w \gg 1$ a violation of the FDT is
observed with $X<1$. The behavior of $X(C)$ for $t/t_w \gg 1$
allows for a simple classification of out-of-equilibrium systems
into three main categories: (\textit{i}) the value $X=0$ is
related to coarsening systems, (\textit{ii}) a constant $X<1$
value is associated with structural glasses, and (\textit{iii}) a
decreasing monotonic $0<X(C)<1$ function is associated to spin
glasses. We will later use this simple scheme to physically
reinterpret the out-of-equilibrium dynamics of spin glasses. In
particular, the continuous line in Fig.~\ref{figure1} shows the
parametric plot for the 3D EA model with linear size $L=20$ for
$t_w=10$. The results correspond to averages over 50 different
samples. After the departure from the quasi-equilibrium regime a
continuous variation of the FDR is observed \cite{Barrat98,Ricci}.

In order to study how the heterogeneities arise in the violation
of the FDT, we measure the joint probability distribution (JPD) of
the single spin quantities, $\rho[{\cal
C}_i(t,t_w),T\chi_i(t,t_w)]$. In Fig.~\ref{figure1} we show the
projection of $\rho$ in the $(T\chi,C)$ plane for the same 50
samples used above. The sequence shows the evolution of the JPD
for increasing $\Delta t= t-t_w$ as a map plot. The parametric
plot (solid curve) of the global quantities is indicated in each
figure as a guide to the eye, and for each value of $\Delta t$ the
mean value of the distribution is indicated with a white circle.

Note that already for very short $\Delta t$ the distribution
presents an elongated shape indicating an heterogeneous behavior.
As $\Delta t$ grows the distribution elongates further and the
presence of two peaks can be clearly observed. At this point, it
is clear that the mean value of the {\em bimodal distribution}
shown in Fig.~\ref{figure1}(d) cannot be used as a representative
quantity to characterize the parametric plot. Since for the
development of the two peaks it is necessary to reach the longest
$\Delta t$ as possible, we chose to work with a low $t_w=10$
value. For longer $t_w$ values we observed the same trend to a
bimodal JPD for long time differences.

In the search for the origin of these {\em strong} dynamical
heterogeneities we focus our analysis on a constrained structure,
which is usually called a backbone. The presence of a backbone in
models defined on randomly connected graphs leads to time-scale
separation and heterogeneous dynamics at sufficiently low
temperatures \cite{Barrat99}. A backbone is also present in the EA
model \cite{Vannimenus,Barahona}, which is formed by those bonds
which are always satisfied or always frustrated in all
ground-state configurations. This structure can be used to divide
the spins in the system in two sets: solidary spins, which are the
spins in the backbone and thus maintain their relative orientation
in all the ground-state configurations, and non-solidary spins
which are simply all the spins that are not in the backbone
\cite{Roma2}. Further insight in the static properties of the 3D
and 2D EA model has been recently addressed by Rom\'a {\em et al.}
\cite{Roma1,Roma3}. They found that the fraction of solidary spins
in 3D(2D) is $0.76$($0.67$) of the total. In contrast with what is
observed in 2D \cite{Barahona,Roma1}, the largest component of the
solidary spins in 3D does not fragment as the system size grows,
involving $60 \%$ of all the spins. Furthermore, the largest
component in 3D forms a compact \textit{percolating cluster}
\cite{Roma3}. We use this information to analyze the strong
dynamical heterogeneities observed in Fig.~\ref{figure1}, and
divide the system in two sets: the spins belonging to the
percolating cluster of solidary spins (set A) and the remaining
ones (set B).

Since we are interested in performing averages over disorder using
this division, a considerable number of different realizations,
where ground-state information is used, need to be taken into
account. In order to calculate the backbone for each sample, we
used an improvement of the Rigid Lattice Searching Algorithm as in
\cite{Roma3}. This requires a large computational effort, since in
order to obtain the exact backbone it is necessary to reach the
ground state $O(N)$ times. This sets a limit to the largest size
we can consider to $N=8^3$ \cite{Roma3}.

\begin{figure}
\includegraphics[width=\linewidth,clip=true]{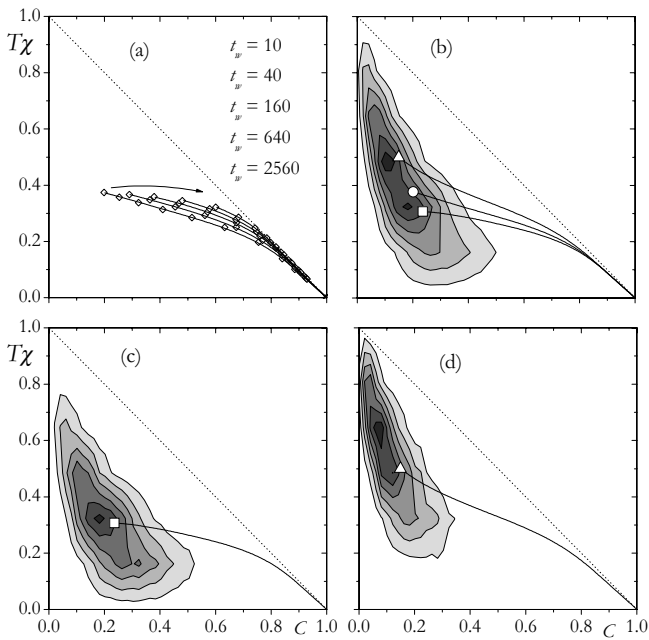}
\caption{\label{figure2} (a) Parametric plot for $L=8$ (continuous
line) and $L=20$ (open diamond) for five different waiting times.
The arrow indicates increasing $t_w$. (b) Projection of $\rho$ and
parametric plot of the global quantities $T\chi(C)$ for $L=8$
(circle). We also present the separation into sets A (square) and
B (triangle). (c) and (d) show $\rho$ and  its corresponding
parametric plot constrained to sets A and B respectively. In (b),
(c) and (d), the white symbols indicate the mean value of the JPD
distribution for $t_w = 10$ and  $\Delta t=10^4$.}
\end{figure}

To establish finite size effects we compare the results obtained
with $L=8$ (over $300$ samples) and $L=20$ (over 50 samples). In
Fig.~\ref{figure2}(a) we show the parametric plot for $L=8$
(continuous line) and $L=20$ (open diamond) for five different
waiting times. For both system sizes the results overlap, which
shows that, up to the times considered, no significant deviations
due to finite size effects are present. Note also the shape of the
JPD for $L=8$ (Fig.~\ref{figure2}(b)) and $L=20$
(Fig.~\ref{figure1}(d)), which shows that for both system sizes
the qualitative results obtained are similar.

Figure~\ref{figure2}(b) also shows the behavior of the global
parametric plot (circle) and its division into sets A (square) and
B (triangle). The JPD constrained to sets A and B are presented in
Figs.~\ref{figure2}(c) and \ref{figure2}(d) respectively. Note
that the mean value of the JPD for the set A (B) is very close to
the lower (upper) peak of the total JPD. In both cases the
distributions are very broad, however the contribution of each set
to the total distribution is clearly divided. Spins in set A (B)
mainly contribute to the lower (upper) peak.

We now present and support a simple physical interpretation to
understand the different behavior of each set of spins. Let us
first highlight  a significant topological property of set A. We
observe that in average approximately  $10 \%$ of the bonds in
this set are frustrated in the ground state. By  using a gauge
transformation for a particular sample, which leaves the
Hamiltonian invariant \cite{Toulouse}, one can change the bonds in
set A so that all its spins are aligned in the ground state. As a
consequence, this set is mapped into a ferromagnet with a low
concentration ($ \approx 10 \%$) of anti-ferromagnetic bonds. Note
that, when the anti-ferromagnetic bond concentration $x$ is
changed in the 3D EA model, it is well established that a
ferromagentic order is present below $x =0.22$ \cite{Hartmann}.
Then, since set A is below this limit ($x \approx 0.1$), we expect
a ferromagnetic-like order to persist in this region up to a
finite temperature.  This result allows us to establish an
unexpected property of the 3D $\pm J$ EA model: \textit{if}
restricted to set A, an out-of-equilibrium behavior in accordance
with that observed in a coarsening system should be found.

\begin{figure}
\includegraphics[width=7cm,clip=true]{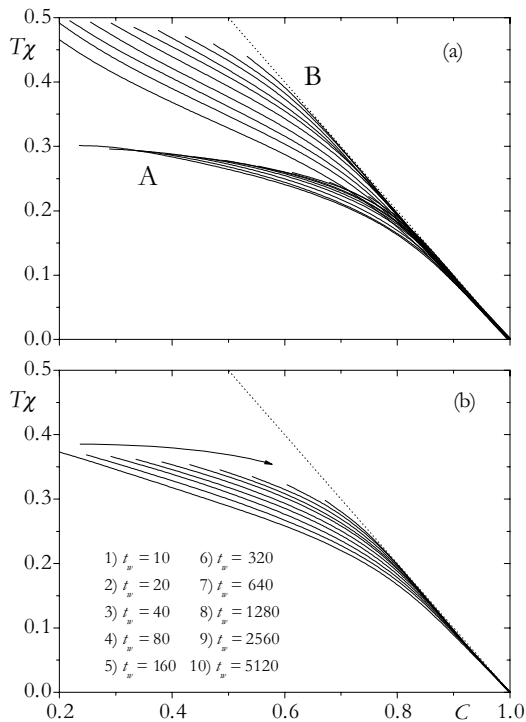}
\caption{\label{figure3} Parametric plot for: (a) sets A and B and
(b) for the system as a whole. The arrow indicates increasing
$t_w$ for the ten different waiting times.}
\end{figure}

In order to analyze the violation of the FDT, we present in
Fig.~\ref{figure3} (a) the parametric plot for each set for ten
different $t_w$. We find that the lower curves, which correspond
to set A, present a clear saturation to a curve away from the FDT
line for increasing $t_w$, which is not observed when the system
is considered as a whole (Fig.~\ref{figure3} (b)). This behavior
reflects the particular topological characteristics of set A. We
expect that for larger system sizes and long times the curves will
follow an out-of-equilibrium behavior similar to the violation of
the FDT in coarsening systems \cite{Crisanti}.

In contrast to the saturation behavior of set A, the upper curves
in Fig.~\ref{figure3} (a) show that set B tends to the FDT limit
for increasing $t_w$. This surprising result is completely
unexpected, since the out-of-equilibrium dynamics of the system is
taking place below the critical temperature $T_c$,  and a
violation of the FDT should be observed when starting from random
initial conditions ($T \gg T_c$) \cite{lesuch1,Crisanti}. Notice
that all the information on the underlying processes, which we
observe using the division into sets A and B, is lost when one
considers the system as a whole (Fig.~\ref{figure3} (b)).

In order to stress the fundamental differences that can be pointed
out in the intuitive frame proposed, we compare the
out-of-equilibrium dynamics of the  EA model in 2D and 3D. These
systems present very similar behavior at low temperatures,
allowing for comparisons between their dynamics
\cite{Roma2,Barrat99,Barrat01}. However, the fact
 that $T_c=0$ in 2D and $T_c>0$ in 3D, recalls that the physical
mechanisms present in the dynamics are clearly different. Notice
also that since there is no percolating cluster of solidary spins
in 2D \cite{Barahona}, the system can only be divided in solidary
and non-solidary spins \cite{Roma2}. Using this division for the
2D EA model we found that, as expected, the evolution of the
parametric plot for both sets tends to the FDT limit for
increasing $t_w$, in contrast to the results obtained in 3D.

Summarizing, we observe strong dynamical heterogeneities in the
JPD of single spin correlation and integrated response functions.
We have gone beyond previous works on single spin quantities, and
have been able to show that these heterogeneities persist even
when averages over disorder are taken into account. Using
information from the ground state topology we are able to
establish a quantitative relation between spatial and dynamical
heterogeneities. In particular, we observe that in the 3D EA model
there is a fraction of the system which is capable of sustaining a
ferromagnetic-like order. The presence of this structure is
reflected in the parametric plot of $T \chi$ vs. $C$. On the other
hand, the other fraction of the system presents a dynamical
behavior that asymptotically tends to satisfy the FDT. As a
consequence, the global out-of-equilibrium dynamics of the system
can be regarded as a combination of two different components. A
comparison of the violation of the FDT in 3D with 2D reveals that
the underlying physical mechanisms present in the dynamics are
clearly different.

We have shown that the continuous violation of the FDT, considered
a signature of full RSB, can be regarded as the combination of two
different components. This particular results suggest that in
short-range spin glasses the  droplet picture could be valid if
restricted to a finite fraction of the system. However, as we have
shown in the 3D EA model, to understand the behavior of the system
as a whole it is also neccesary to take into account the existence
of a finite fraction of the system which presents a paramagnetic
behavior. We believe that the inclusion of these elements in a new
theoretical description can contribute to the understanding of the
physical nature of spin glasses. In particular it can contribute
to the study of the relation between the percolation problem and
the spin-glass transition \cite{Stauffer}, and also to hot topics
such as temperature and disorder chaos \cite{Katz2}.

We thank  L.F. Cugliandolo, J. Kurchan and S. Risau-Gusman for
fruitful discussions. We acknowledge financial support from CNEA
and CONICET (Argentina), and ICTP NET-61 (Italy). F.R., S.B and D.
D acknowledge support from CONICET, PIP05-5596. F.R. thanks Univ.
Nac. de San Luis (Argentina) project 322000 and Millennium
Scientific Iniciative (Chile) contract P-02-054-F for partial
support. P.M.G. acknowledges financial support from CONICET
PIP05-5114 (Argentina) and ANPCyT PICT03-13893 (Argentina).

\end{document}